\documentclass{article}

\usepackage[english]{babel}
\usepackage{csquotes}

\usepackage[letterpaper,top=2cm,bottom=2cm,left=3cm,right=3cm,marginparwidth=1.75cm]{geometry}

\usepackage{amsmath}
\usepackage{graphicx}
\usepackage[colorlinks=true, allcolors=blue]{hyperref}
\usepackage{float}
\usepackage{biblatex}
\usepackage{sidecap}
\sidecaptionvpos{figure}{c}

\title{Volumetric and structural connectivity abnormalities co-localise in TLE}
\author{Jonathan J. Horsley$^{1}$, Gabrielle M. Schroeder$^{1}$, \\Rhys H. Thomas$^{2}$, Jane de Tisi$^{3}$, Sjoerd B. Vos$^{3,4,5}$, Gavin P. Winston$^{3,6}$, \\John S. Duncan$^{3}$, Yujiang Wang$^{1,2,3}$ and Peter N. Taylor$^{*1,2,3}$}

\date{\today}
\begin{document}
\maketitle

\begin{enumerate}
\item{CNNP Lab (www.cnnp-lab.com), Interdisciplinary Computing and Complex BioSystems Group, School of Computing, Newcastle University, Newcastle upon Tyne, United Kingdom}
\item{Faculty of Medical Sciences, Newcastle University, Newcastle upon Tyne, United Kingdom}
\item{Neuroradiological Academic Unit, UCL Queen Square Institute of Neurology, University College London, London, United Kingdom}
\item{Centre for Microscopy, Characterisation, and Analysis, The University of Western Australia, Nedlands, Australia}
\item{Centre for Medical Image Computing, Computer Science Department, University College London, London, United Kingdom }
\item{Division of Neurology, Department of Medicine, Queen's University, Kingston, Canada}
\end{enumerate}
\begin{center}
* Peter.Taylor@newcastle.ac.uk
\end{center}

\begin{abstract}
Patients with temporal lobe epilepsy (TLE) exhibit both volumetric and structural connectivity abnormalities relative to healthy controls. How these abnormalities inter-relate and their mechanisms are unclear. We computed grey matter volumetric changes and white matter structural connectivity abnormalities in 144 patients with unilateral TLE and 96 healthy controls. Regional volumes were calculated using T1-weighted MRI, while structural connectivity was derived using white matter fibre tractography from diffusion-weighted MRI. For each regional volume and each connection strength, we calculated the effect size between patient and control groups in a group-level analysis. We then applied hierarchical regression to investigate the relationship between volumetric and structural connectivity abnormalities in individuals. Additionally, we quantified whether abnormalities co-localised within individual patients by computing Dice similarity scores. \\

In TLE, white matter connectivity abnormalities were greater when joining two grey matter regions with abnormal volumes. Similarly, grey matter volumetric abnormalities were greater when joined by abnormal white matter connections. The extent of volumetric and connectivity abnormalities related to epilepsy duration, but co-localisation did not. Co-localisation was primarily driven by neighbouring abnormalities in the ipsilateral hemisphere. Overall, volumetric and structural connectivity abnormalities were related in TLE. Our results suggest that shared mechanisms may underlie changes in both volume and connectivity alterations in patients with TLE.
\end{abstract}

\section{Introduction}

Epilepsy is a common neurological disorder affecting around 50 million people worldwide \cite{fiest_prevalence_2016}. There are many different types of epilepsy, but the most common are focal epilepsies and in particular temporal lobe epilepsy (TLE). In TLE, a variety of structural alterations have been found in grey and white matter \cite{whelan_structural_2018} \cite{hatton_white_2020}. Both grey and white matter alterations may be related to the causes and consequences of the disorder, but it is currently unclear how these changes are inter-related within individual patients. Improving our knowledge of the relationship between volumetric measures of grey matter and structural connectivity in white matter is important for understanding the pathophysiology of TLE. \\

TLE has been considered a focal disorder characterised with grey matter lesions and/or volume atrophy. The hippocampus in particular is a key epileptogenic region \cite{engel_mesial_2001}, with atrophy observed in patients with hippocampal sclerosis \cite{thom_review_2014}. Numerous other studies have observed significant volume reductions in the ipsilateral hippocampus compared to healthy controls \cite{whelan_structural_2018} \cite{keller_morphometric_2015} \cite{farid_temporal_2012} \cite{winston_automated_2013}. However, atrophy is not restricted to the hippocampus and is also seen in ipsilateral subcortical and temporal regions, including the entorhinal cortex \cite{bonilha_medial_2003} \cite{keller_morphometric_2015}, thalamus \cite{whelan_structural_2018}, temporal gyri, parahippocampal gyrus, and temporal pole \cite{moran_extrahippocampal_2001}. Moran et al. (2001) found a positive correlation between the severity of hippocampal and extrahippocampal atrophy, suggesting a common mechanism for abnormalities across different brain regions \cite{moran_extrahippocampal_2001}. Some studies have also identified more widespread atrophy, including in the frontal lobe \cite{doucet_frontal_2015}, and in contralateral temporal and subcortical regions \cite{araujo_volumetric_2006} \cite{keller_morphometric_2015}. Both cross-sectional \cite{bernasconi_progression_2005} and longitudinal data \cite{alvim_progression_2016} have related the degree of atrophy to epilepsy duration. Additionally, volumetric abnormalities can lateralise the side of seizure onset with a high degree of accuracy \cite{li_lateralization_2000}\cite{farid_temporal_2012}. Overall, these findings demonstrate the prevalence of volumetric abnormalities in TLE, particularly in ipsilateral subcortical and temporal regions, and their relation to clinical factors such as epilepsy duration. \\

Neuroimaging techniques such as diffusion MRI tractography allow non-invasive modelling of the connections between different brain regions \cite{basser_vivo_2000}. Fractional anisotropy (FA) quantifies the degree to which diffusion of water molecules in white matter axons is directional. FA is often used as a marker for axonal integrity and structural connectivity in epilepsy \cite{van_diessen_functional_2013}. Distinct regions of the brain can be modelled as nodes and the connections between them as edges \cite{bassett_network_2017}. This allows network analysis and the facilitates the study of epilepsy as a network disorder \cite{bernhardt_network_2015} \cite{spencer_neural_2002}. Studies of diffusion data have examined structural connectivity in TLE. Compared to healthy controls, patients with TLE exhibit reduced FA in both the ipsilateral anterior and mesial temporal lobe \cite{riley_altered_2010}, reduced FA in the ipsilateral parahippocampal cingulum and external capsule \cite{hatton_white_2020}, and reduced connectivity between the ipsilateral thalamus and precentral gyrus \cite{bonilha_presurgical_2013}. Besson et al. (2014) found that patients with left TLE had FA reductions that were strongly lateralised to the ipsilateral temporal lobe, whilst FA reductions in patients with right TLE were less extreme and restricted to bilateral limbic and ipsilateral temporal cortex \cite{besson_structural_2014}. However, structural connectivity may also be significantly decreased beyond epileptogenic regions \cite{besson_anatomic_2017}. These structural connectivity changes may relate to duration \cite{owen_multivariate_2021} \cite{hatton_white_2020}, and be predictive of surgical outcome \cite{taylor_impact_2018} \cite{sinha_structural_2020} \cite{bonilha_brain_2015} or secondary generalisation of seizures \cite{sinha_focal_2021}. \\

Only few studies have considered both volumetric and structural connectivity abnormalities. Keller et al. (2012) found associations between FA reductions within regions and volume atrophy in a subset of subcortical regions, including the contralateral hippocampus and bilateral thalami \cite{keller_concomitant_2012}. Atrophy of the hippocampus in patients with hippocampal sclerosis has also been related to whole-brain alterations in network properties including path length and clustering \cite{bernhardt_hippocampal_2019}. Another study found that grey matter atrophy co-localised with hub regions inferred from the connectome of healthy subjects, but did not consider connectivity abnormalities \cite{lariviere_network-based_2020}. To our knowledge, no study has investigated the relationship between abnormalities in the two modalities by considering regions and connections in a within-patient analysis. \\

It is currently unknown whether volumetric and connection abnormalities co-localise within individual patients and whether they have shared mechanisms that drive their co-localisation. Clearly, both volume and connectivity provide useful, and potentially complementary, information and should be considered simultaneously to to better understand the pathophysiology of TLE. Importantly, we move beyond group-level analysis and perform hierarchical statistical modelling to quantify abnormality co-occurrence within \textit{individual} patients. Within-patient analysis allows for individual patient models and predictions, and allows for the study of disease progression in a cross-sectional manner. \\

\begin{SCfigure}
\centering
\includegraphics[width=0.5\textwidth]{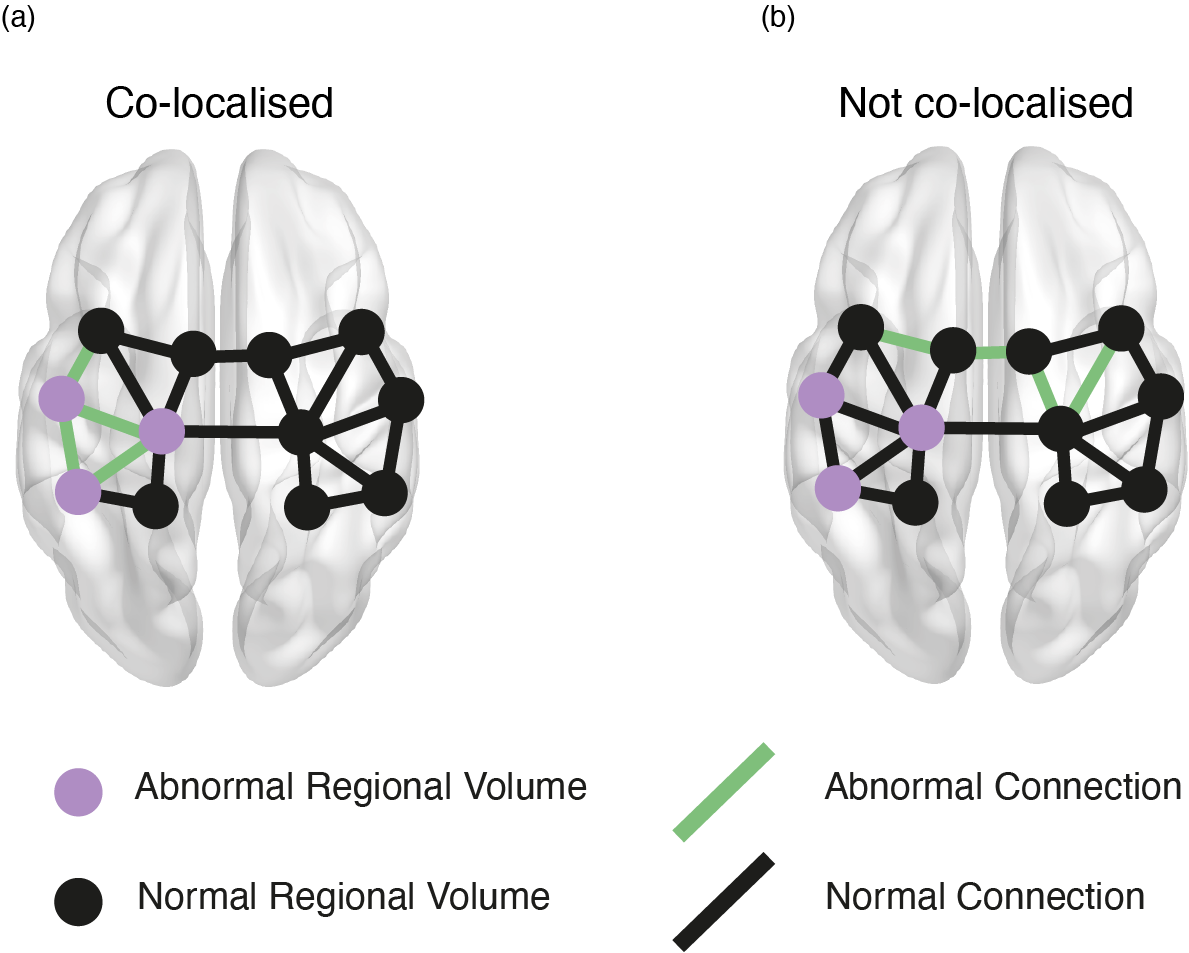}
\caption{\label{fig:Schematic}\textbf{Possible relationships between volumetric and connectivity abnormalities in TLE.} a) Volumetric and connectivity abnormalities are co-localised if nodes with abnormal volumes also have abnormal connections. b) Volume and connectivity abnormalities are not co-localised if nodes with abnormal volumes have normal connections, and abnormal connections are connected to normal volumes.}
\end{SCfigure}

In this study, we analyse the relationship between volumetric and structural connectivity abnormalities to address the following questions:

\begin{enumerate}
  \item At a group level, where do volumetric and connectivity abnormalities occur?
  \item Within \textit{individual} patients, are connectivity abnormalities greater when joining regions with abnormal volumes? Similarly, are volumetric abnormalities greater when joined by abnormal connections? 
  \item Within \textit{individual} patients, do regions (network nodes) with atrophy also have abnormal connections (co-localise) or not (Figure \ref{fig:Schematic}), and does co-localisation relate to epilepsy duration?
\end{enumerate}

\section{Methods}

\subsection{Patients and MRI acquisition}

We retrospectively studied two cohorts of patients with unilateral TLE from the National Hospital of Neurology and Neurosurgery, London, United Kingdom. Pseudonymised data were analysed under the approval of the Newcastle University Ethics Committee (2225/2017). Both cohorts had an accompanying sample of healthy controls. All subjects underwent anatomical T1-weighted MRI and diffusion-weighted MRI. In total, there were 144 patients and 96 healthy controls. \\

The first cohort consisted of 84 patients and 29 healthy controls. For this cohort, MRI studies were performed on a 3T GE Signa HDx scanner (General Electric, Waukesha, Milwaukee, WI). Standard imaging gradients with a maximum strength of $40mT m^{-1}$ and slew rate $150T m^{-1} s^{-1}$ were used. All data were acquired using a body coil for transmission, and 8-channel phased array coil for reception. Standard clinical sequences were performed including a coronal T1-weighted volumetric acquisition with 170 contiguous 1.1 mm-thick slices (matrix, 256 × 256; in-plane resolution, 0.9375 × 0.9375 mm). \\

Diffusion-weighted MRI data were acquired using a cardiac-triggered single-shot spin-echo planar imaging sequence \cite{wheeler-kingshott_investigating_2002} with echo time = 73$ms$. Sets of 60 contiguous 2.4 mm-thick axial slices were obtained covering the whole brain, with diffusion sensitizing gradients applied in each of 52 noncollinear directions (b value of 1,200 $s/mm^{2}$ [$\delta = 21 ms$, $\Delta = 29 ms$, using full gradient strength of 40$mT m^{-1}$]) along with 6 non-diffusion-weighted scans. The gradient directions were calculated and ordered as described elsewhere \cite{cook_optimal_2007}. The field of view was 24$cm$, and the acquisition matrix size was 96 × 96, zero filled to 128 × 128 during reconstruction, giving a reconstructed voxel size of 1.875 × 1.875 × 2.4 mm. The DTI acquisition time for a total of 3480 image slices was approximately 25 min (depending on subject heart rate). \\

The second cohort consisted of 60 patients and 67 healthy controls. For this cohort, MRI studies were performed on a 3T GE MR750 scanner. Standard imaging gradients with a maximum strength of 50$mT m^{-1}$ and slew rate 200$T m^{-1} s^{-1}$ were used. All data were acquired using a body coil for transmission, and 32-channel phased array coil for reception. Standard clinical sequences were performed including a coronal T1-weighted volumetric acquisition with 224 contiguous 1mm-thick slices (matrix, 256 × 256; in-plane resolution, 1 × 1 mm). \\

Diffusion-weighted MRI data were acquired using a single-shot spin-echo planar imaging sequence with echo time = 74.1$ms$. Sets of 70 contiguous 2 mm-thick axial slices were obtained covering the whole brain. A total of 115 volumes were acquired with 11, 8, 32, and 64 gradient directions at b-values of 0, 300, 700, and 2500 $s/mm^{2}$ respectively ($\delta = 21.5 ms$, $\Delta = 35.9 ms$) as well as a single b = 0-image with reverse phase-encoding (B0). The field of view was 25.6$cm$, and the acquisition matrix size was 128 × 128, giving a reconstructed voxel size of 2 × 2 x 2 mm.

\subsection{Image processing}

\subsubsection{T1 processing}

T1-weighted MRI was used to generate parcellated grey matter regions of interest. FreeSurfer's recon-all pipeline was applied (https://surfer.nmr.mgh.harvard.edu/), which performs intensity normalization, skull stripping, subcortical volume generation, gray/white segmentation, and parcellation \cite{fischl_freesurfer_2012}. Surfaces and volumes were corrected where appropriate according to ENIGMA pipelines \cite{hatton_white_2020}\cite{whelan_structural_2018}. The default parcellation scheme from FreeSurfer (the Desikan-Killiany atlas \cite{desikan_automated_2006}) contains 82 cortical and subcortical regions and is widely used in the literature \cite{munsell_evaluation_2015}\cite{taylor_structural_2015}. Additionally, further denominations of the Desikan-Killiany atlas were used, with 128, 233 and 462 regions to investigate consistency across parcellations.

\subsubsection{DWI processing}

Diffusion-weighted MRI data were first corrected for signal drift \cite{vos_importance_2017}, then eddy current and movement artefacts were corrected using the FSL eddy\_correct tool \cite{andersson_integrated_2016} (first cohort) or using EDDY/TOPUP (second cohort). The b vectors were then rotated appropriately using the ‘fdt-rotate-bvecs’ tool as part of FSL \cite{jenkinson_fsl_2012}\cite{leemans_b_2009}. The diffusion data were reconstructed in MNI-152 space using q-space diffeomorphic reconstruction (QSDR) \cite{yeh_ntu-90_2011} with a diffusion sampling length ratio of 1.2. The HCP-1065 tractography atlas \cite{yeh_population-averaged_2018} was used to determine connections between regions. The use of a tractography atlas is expected to result in fewer false positive connections than fibre tracking algorithms, since each tract has been visually confirmed to be expected and not spurious. This approach has the benefit of reducing the influence of network density on the subsequent group comparisons of networks \cite{van_wijk_comparing_2010}, and has been used previously \cite{sinha_focal_2021} \cite{moreira_da_silva_investigating_2020}. A connection between MNI-152 space regions of the same parcellation was defined as present if streamlines passed into both regions in the corresponding region pair.

\subsection{Data processing}

\begin{figure}[H]
\centering
\includegraphics[width=1\textwidth]{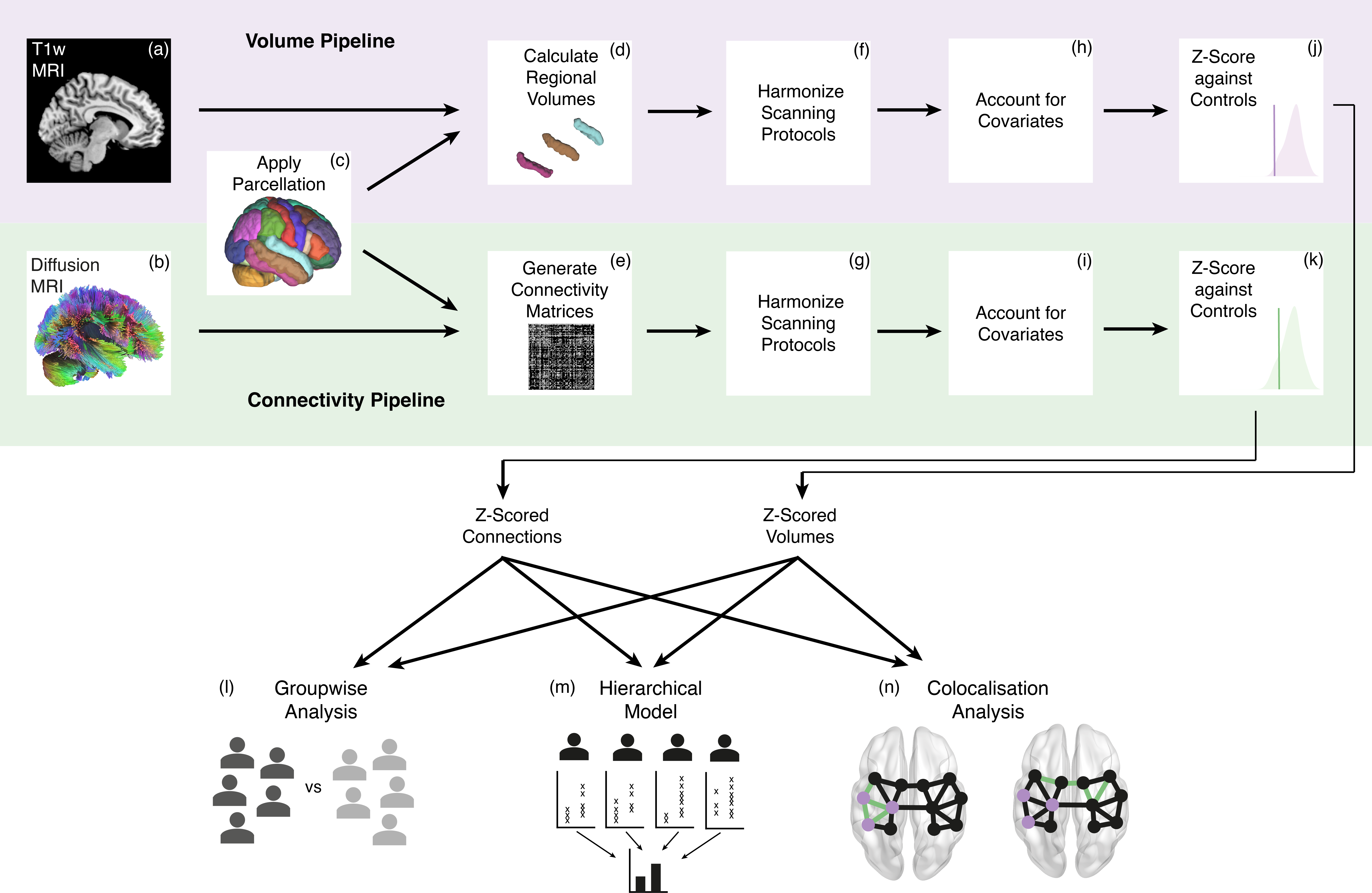}
\caption{\label{fig:Methods_Fig}\textbf{Processing and analysis pipeline.} We used T1-weighted MRI (a) with the Desikan-Killiany parcellation scheme (c) to calculate regional volumes for each subject (d). We also used diffusion-weighted MRI (b) with the same parcellation scheme (c) to generate connectivity matrices using average FA between regions (e). Both regional volumes (f) and connections (g) were harmonized across scanning protocols using ComBat. Age and sex were accounted for in both pipelines, with ICV also accounted for in the volume pipeline (h, i). Within subjects, each regional volume (j) and connection strength (k) were expressed as $z$-scores relative to a cohort of healthy controls. These $z$-scores formed the input for our group-level analysis (l), hierarchical models (m) and co-localisation analysis (n).}
\end{figure}

All data processing was performed using R version 4.02 (https://www.r-project.org), unless otherwise stated. The overall processing and analysis pipeline is shown in Figure \ref{fig:Methods_Fig}. 

\subsubsection{Volume}

The volume of each region in each subject was computed using FreeSurfer. The two scanning protocols were systematically different; therefore, ComBat was applied to remove scanner differences while preserving biological variability \cite{fortin_harmonization_2017}. Accounting for known covariates (age, sex and ICV) using robust linear regression applied to healthy controls, we calculated volume residuals of each region in each subject. These volume residuals captured how much a regional volume differed to the average healthy control given a subject's age, sex and ICV. We then transformed the residuals into $z$-scores based on the volume distribution of healthy controls.

\subsubsection{Connectivity}

Connectivity matrices were computed for each subject using DSI Studio (http://dsi-studio.labsolver.org), using the average FA of connections between regions as a measure of connectivity strength. In a similar way to volume, ComBat was applied to each connection to remove scanner differences. Known covariates (age and sex) were then accounted for using robust linear regression, which was applied to healthy controls to calculate connection residuals. These connection residuals quantify the amount by which the average FA in a connection differs compared to the average healthy control given a subject's age and sex. We then transformed the residuals into $z$-scores based on the connection distribution of healthy controls.

\subsection{Statistical analysis}

We flipped the z-scores of right and left hemispheres in patients with RTLE to perform an ipsilateral-contralateral analyses. This flipping was only done after each region and connection was z-scored relative to controls. This approach ensured that region and connection abnormalities were only calculated by comparison against the same hemisphere in controls. This comparison is important since left and right hemispheres are not simply mirror images of one another. Our approach preserved our large sample size and followed previous literature \cite{taylor_impact_2018} \cite{sinha_structural_2020} \cite{keller_morphometric_2015}.

\subsubsection{Group-level analysis}

Using the volume $z$-scores of each region (82 in total), we computed an effect size between patients and healthy controls using Cohen's $d$, defined as:

\begin{align*}
\text{Cohen's $d$} &= \frac{\overline{x_{P}} - \overline{x_{C}}}{s},
\end{align*}

where $\overline{x_{P}}$ is the mean volume $z$-score of a region in patients, $\overline{x_{C}}$ is the mean volume $z$-score of a region in controls and $s$ is the pooled standard deviation. \\

Similarly, using the connection $z$-scores between each pair of regions (1289 in total), we computed an effect size between patients and healthy controls using Cohen's d.

\subsubsection{Hierarchical modelling}

Methods in the previous section examined volumetric and connectivity abnormalities at a group level, but did not provide information at an individual patient level. However, group-level analysis has the potential to be misleading. As an example, consider a cohort with 50\% of patients with volumetric, but not connectivity, abnormalities in ipsilateral temporal regions (Supplementary Analysis 1 - Figure S1a). If the other 50\% of patients had connectivity, but not volumetric, abnormalities in ipsilateral temporal regions (Figure S1b) then, at a group level, the abnormalities would appear to co-localise. However, in this example scenario, no single patient would have \textit{both} volumetric and connectivity abnormalities together in ipsilateral temporal regions. A traditional group analysis is therefore insensitive to within-patient features. In contrast, hierarchical modelling uses within-patient information to investigate the relationship between volumetric and connectivity abnormalities; one of the key novelties of our study. \\

In our cohort of 144 patients, volumetric and connection abnormalities were potentially inter-related on a subject-specific level. To account for this nested nature of the data, we fitted a two-level hierarchical model to our data, allowing random intercepts and slopes between individual subjects. Hierarchical modelling is appropriate for nested data organized at more than one level (e.g. many connections within individual patients) and accounts for a potential lack of independence between connection $z$-scores within the same patient \cite{osborne_advantages_2000}. For example, the probability of reduced FA in a connection is likely related to the quantity and location of other FA reductions in that patient. The hierarchical modelling approach also allows for individual heterogeneity to inform the overall model estimates. \\

Abnormal regions were defined as those with volumes below the 2.5th percentile threshold ($z < -1.96$). This threshold value was scanned from $-1.0$ to $-2.5$ in steps of $0.1$ to determine the robustness of results. Specifically, we modelled whether connection abnormalities ($z$-scores) were larger when they connected one or two volumetric abnormalities, compared to the case when they connected two normal ROIs. The hierarchical model was fitted using restricted maximum likelihood (REML) from the the `lmer' function in R's `lme4' package. \\

Mathematically, we modelled the connection abnormality $C_{ijk}$ for the connection between regions $i$ and $j$ in subject $k$ with the following system of equations: \\

Level One:
\begin{align*}
C_{ijk} &= a_{k} + b_{k}V_{ijk}^\prime + c_{k}V_{ijk}^{\prime\prime} + \epsilon_{ijk}
\end{align*}

Level Two:
\begin{align*}
a_{k} &= \alpha_{0} + u_{k} \\
b_{k} &= \beta_{0} + v_{k} \\
c_{k} &= \gamma_{0} + w_{k},
\end{align*}

\begin{table}[ht]
\centering
\begin{tabular}{c | p{0.85\linewidth}}
 \hline
 Coefficient & Interpretation \\ [0.5ex] 
 \hline\hline
 $C_{ijk}$ & the connection $z$-score for the connection between regions $i$ and $j$ in subject $k$ \\ [0.5ex] 
 $a_{k}$ & mean connection $z$-score in subject $k$ for connections between two normal regions \\ [0.5ex] 
 $b_{k}$ & mean change in connection $z$-score in subject $k$ for connections between one normal and one abnormal region, as compared to two normal regions \\ [0.5ex] 
 $c_{k}$ & mean change in connection $z$-score in subject $k$ for connections between two abnormal regions, as compared to two normal regions \\ [0.5ex] 
 $\epsilon_{ijk}$ & difference between fitted and observed abnormalities for the connection between regions $i$ and $j$ in subject $k$ \\ [0.5ex] 
 $V_{ijk}^\prime$ & $\begin{cases}
                1, & \text{if connection is between one normal and one abnormal region}\\
                0, & \text{otherwise}
            \end{cases}$ \\ [1ex]
 $V_{ijk}^{\prime\prime}$ & $\begin{cases}
                1, & \text{if connection is between two abnormal regions}\\
                0, & \text{otherwise}
            \end{cases}$ \\ [1ex]
 $\alpha_{0}$ & true mean connection $z$-score for connections between two normal regions across all subjects \\ [0.5ex] 
 $u_{k}$ & difference between $\alpha_{0}$ and mean connection $z$-score for connections between two normal regions in subject $k$ \\ [0.5ex] 
 $\beta_{0}$ & true mean difference in connection $z$-score for connections between one normal and one abnormal region, as compared to two normal regions across all patients \\ [0.5ex] 
 $v_{k}$ & difference between $\beta_{0}$ and mean difference in connection $z$-score for connections between one normal and one abnormal region, as compared to two normal regions in subject $k$ \\ [0.5ex]
 $\gamma_{0}$ & true mean difference in connection $z$-score for connections between two abnormal regions, as compared to two normal regions across all patients \\ [0.5ex] 
 $w_{k}$ & difference between $\gamma_{0}$ and mean difference in connection $z$-score for connections between two abnormal regions, as compared to two normal regions in subject $k$ \\
 \hline
\end{tabular}
\caption{Hierarchical model coefficient interpretations.}
\label{table:1}
\end{table}

The interpretation of the model coefficients are given in Table 1. $V_{ijk}^{\prime}$ and $V_{ijk}^{\prime\prime}$ capture whether a connection in subject $k$ is between one or two abnormal regions, respectively. In this model, there are three key fixed effects to estimate: $\alpha_{0}$, $\beta_{0}$ and $\gamma_{0}$, which allow us to test our hypotheses. The error terms $\epsilon_{ijk}$, $u_{k}$, $v_{k}$ and $w_{k}$ represent random effects. These error terms describe specific patients, which are a sample of a larger population (e.g. our cohort as a sample of all patients with temporal lobe epilepsy). We account for the influence of these random effects in our model, but do not draw conclusions about specific levels. \\

For comparison, and as a null model, we also fit our hierarchical model to two other cases. First, we shuffled the connection $z$-scores and re-fit the same model. Second, we fit the model to our cohort of healthy controls. \\

We had two main hierarchical models. First, we modelled whether volumetric abnormalities coincided with adjacent connectivity abnormalities. Second, we investigated if connection abnormalities coincided with adjacent volumetric abnormalities. Abnormal connections were defined as those with FA reductions below the 2.5th percentile threshold ($z < -1.96$). Again, this threshold value was scanned (-1.0 to -2.5 in steps of 0.1) to determine the robustness of results (see Supplementary Analysis 4). Specifically, we modelled whether the mean volume abnormality of connected regions was larger when the regions were connected by abnormal connections. This is a near-identical approach as to the first hierarchical model, and is fully detailed in Supplementary Analysis 3 for completeness.

\subsubsection{Co-localisation analysis}

Finally, we investigated if abnormalities co-localised (occurred in connected regions) more than would be expected by chance. \\

We first defined a measure of co-localisation. Taking each connection $z$-score, along with the volume $z$-scores of the two connected regions, we applied a threshold in a similar way as to the previous section. Volumes and connections with $z$-scores below the 5th percentile ($z < -1.645$) were deemed abnormal. Z < -1.645 defines the lower 5\% of a standard normal distribution, we see no deviation from normality in our population of controls. This slightly less stringent threshold was used to ensure that a greater proportion of patients had abnormalities in both modalities. This threshold value was scanned to ensure robust results, and these results are shown in Supplementary Analysis 5.\\

\begin{table}[H]
\centering
\begin{tabular}{c|c|c|c|c}
 \hline
 Region $i$ & Region $j$ & $V_{ik}^\prime$ & $C_{ijk}^\prime$ & $V_{jk}^\prime$ \\ [0.5ex]
 \hline\hline
 Ipsilateral Thalamus & Ipsilateral Hippocampus & 0 & 0 & 1 \\ [0.5ex] 
 Ipsilateral Thalamus & Ipsilateral Amygdala & 0 & 1 & 0 \\ [0.5ex] 
 Ipsilateral Thalamus & Ipsilateral Temporal Pole & 0 & 1 & 1 \\ [0.5ex] 
 ... & ... & ... & ... & ... \\
 Contralateral Insula & Contralateral Transverse Temporal Gyrus & 0 & 0 & 0 \\
 \hline
\end{tabular}
\caption{\textbf{Volumetric and connectivity abnormalities in example subject \textit{k}}. In columns $V_{ik}^\prime$, $C_{ijk}^\prime$ and $V_{jk}^\prime$, 0 denotes normality and 1 denotes abnormality. In this subject, both the ipsilateral hippocampus and ipsilateral temporal pole have abnormal volumes. The connections ipsilateral thalamus - ipsilateral amygdala and ipsilateral thalamus - ipsilateral temporal pole are abnormal. Using the Desikan-Killiany atlas with the HPC1065 tractography atlas, there were 1289 unique connections between regions.}
\label{table:2}
\end{table}

To gauge the similarity of the spatial locations of our abnormalities within a patient, we computed the Dice similarity of the volumetric abnormalities and joining connection abnormalities. For each unique connection, we first nominally assigned an \textit{originating} and \textit{destination} region. Then $\pmb{V}_{ik}^\prime$ represented the set of \textit{originating} volumetric abnormalities within subject k (column 3 in Table \ref{table:2}), $\pmb{C}_{ijk}^\prime$ represented the set of connectivity abnormalities within subject k (column 4 in Table \ref{table:2}), and $\pmb{V}_{jk}^\prime$ represented the set of \textit{destination} volumetric abnormalities within subject k (column 5 in Table \ref{table:2}). Then:

\begin{align*}
DS_{ik} = \frac{|\pmb{V}_{ik}^\prime \cap \pmb{C}_{ijk}^\prime|}{|\pmb{V}_{ik}^\prime| + |\pmb{C}_{ijk}^\prime|} && DS_{jk} = \frac{|\pmb{C}_{ijk}^\prime \cap \pmb{V}_{jk}^\prime|}{|\pmb{C}_{ijk}^\prime| + |\pmb{V}_{jk}^\prime|}
\end{align*}

\begin{align*}
DS_{k} &= \frac{DS_{ik} + DS_{jk}}{2}
\end{align*}

In summary, the Dice similarity quantified the extent to which abnormalities occurred simultaneously in white matter connections and the connected grey matter regions, producing a single value, $DS_{k}$, per patient. Because $DS_{k}$ was biased by the number of abnormalities of each set, a greater number of abnormalities generally produced a greater dice similarity score. As a result, we randomly permuted the connection abnormalities and recomputed Dice similarity 5,000 times and defined our unbiased co-localisation measure as:

\begin{align*}
\text{Co-localisation} &= \frac{1}{5000} \sum_{n=1}^{5000} \begin{cases}
                1, & DS_{\text{actual}} > DS_{\text{n}} \\
                0, & DS_{\text{actual}} \leq DS_{\text{n}}
            \end{cases}.
\end{align*}

Co-localisation was a number between 0 and 1 for each patient, unbiased by the number of abnormalities, where 1 signifies that abnormalities occurred in adjacent regions/connections above chance and 0 signifies that abnormalities occurred in adjacent regions/connections below chance. \\

It is plausible that the relationship between volumetric and connectivity abnormalities is related to disease progression. Over time, atrophy may spread from region to region via abnormal connections. To investigate this, we used Pearson correlation to relate epilepsy duration with our co-localisation measure, in addition to the proportion of abnormal regional volumes and proportion of abnormal connections. \\

Finally, in those patients with abnormalities co-localising across the full brain (co-localisation $>$ 0.95), we re-computed co-localisation separately for ipsilateral-only and contralateral-only connections. We performed a paired Mann-Whitney U test to determine whether abnormalities were more likely to co-localise in the ipsilateral hemisphere compared to the contralateral hemisphere. \\

\section{Results}

The results are presented in four sections. First, we highlight abnormalities in volume and connectivity in a group-level analysis. Second, we show how volumetric abnormalities relate to adjacent connectivity abnormalities within individuals using hierarchical modelling. Third, we show how connectivity abnormalities relate to adjacent volumetric abnormalities within individuals using hierarchical modelling. Finally, we investigate whether adjacent volumetric and connectivity abnormalities within individual patients relate to epilepsy duration, surgical outcome and secondary generalization of seizures.

\subsection{Volumetric and connectivity abnormalities coexist in patients}

\begin{figure}[H]
\centering
\includegraphics[width=1\textwidth]{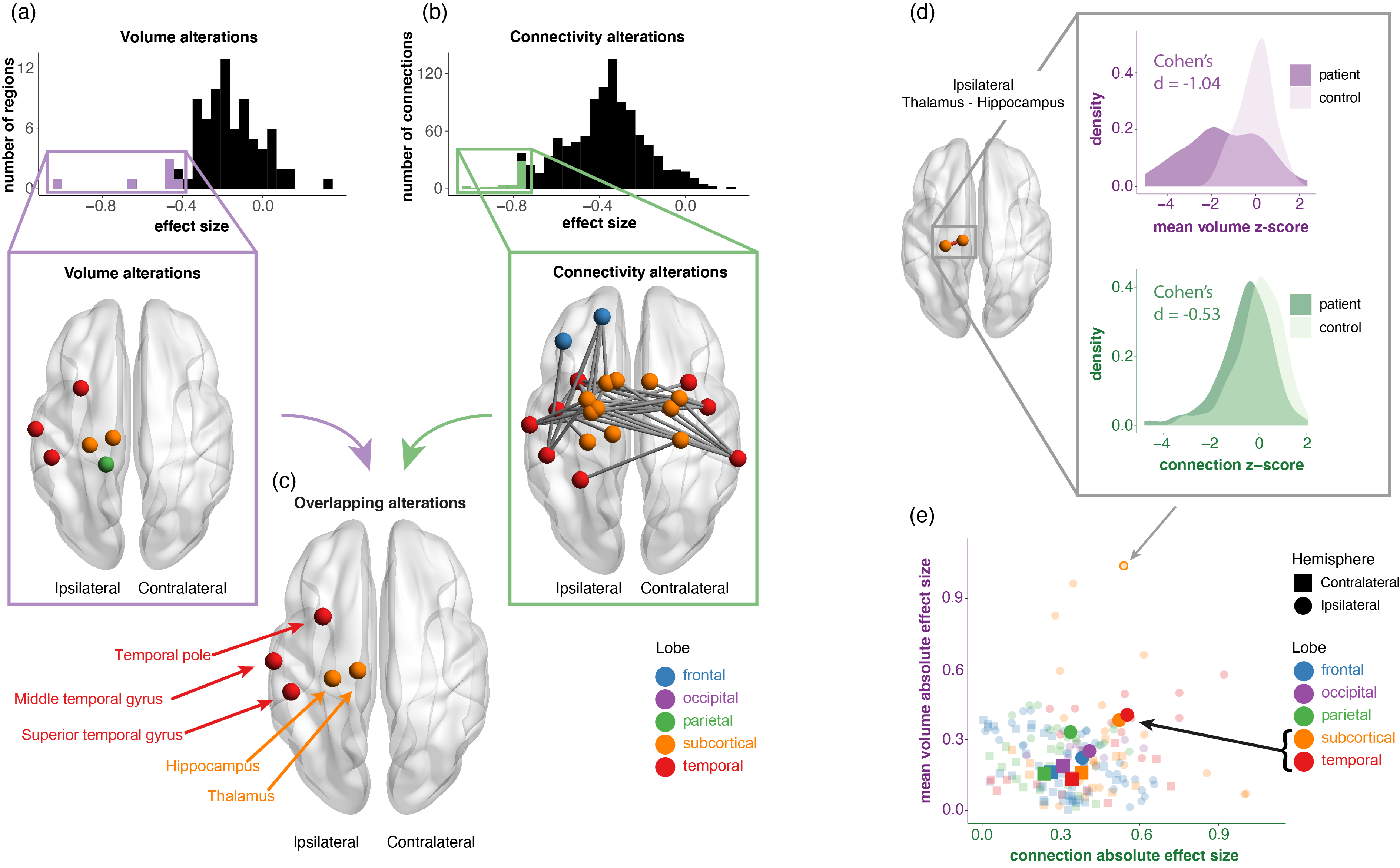}
\caption{\label{fig:Fig1}\textbf{Volumetric and connectivity abnormalities coexist in patients.} At a group level, a) the distribution and location of regional volumetric abnormalities in patients compared to controls. Negative effect sizes indicate a reduction in patients. b) The distribution and location of structural connectivity abnormalities in patients compared to controls. The colour of the node denotes the lobe. c) Regions identified as having both reduced volume and connectivity were located exclusively in ipsilateral subcortical and temporal areas. d) For one specific connection (ipsilateral thalamus - ipsilateral hippocampus), both the effect size of the connection abnormality and the effect size of the mean volumetric abnormality is shown. e) More generally, the relationship between the two modalities are shown for within-lobe connections. Specific connections are represented by small, faint points, and the lobe means are in large, bold points.
}
\end{figure}

For the volume of each region, we computed an effect size between patients and healthy controls (Figure \ref{fig:Fig1}a). We observed reduced volumes in several regions, which were primarily in ipsilateral temporal and subcortical areas, most notably the ipsilateral hippocampus (Cohen's $d$ = -1.02) and ipsilateral thalamus ($d$ = -0.65). \\

Similarly, for the FA of each connection, we computed an effect size between patients and healthy controls (Figure \ref{fig:Fig1}b). We observed widespread structural connectivity reductions in patients relative to controls. Connections between temporal and subcortical regions bilaterally, and ipsilateral frontal regions, especially those involving the ipsilateral temporal pole were most affected. \\

At a group level, regions with the largest volumetric atrophy and structural connectivity reductions (in white matter) were exclusively located in ipsilateral temporal and subcortical areas. These regions were the thalamus, hippocampus, superior temporal gyrus, middle temporal gyrus and temporal pole (Figure \ref{fig:Fig1}c). A full list of regions and their associated lobe is given in Supplementary Analysis 7. \\

Next, we related within-lobe volumetric abnormalities to connectivity abnormalities at a group level. We hypothesised that larger connection abnormalities would coincide with larger neighbouring volumetric abnormalities. For each connection within each subject, we calculated the mean of the volume $z$-scores of the two connected regions, and calculated effect sizes for the connection $z$-score and the mean volume $z$-score between patients and healthy controls. While there was limited evidence for our hypothesis at the connection-level, lobes with larger connection abnormalities displayed larger volumetric abnormalities (Figure \ref{fig:Fig1}e). Ipsilateral temporal and subcortical lobes were more abnormal than other areas (see inset black arrow in Figure \ref{fig:Fig1}e).

\subsection{Connection abnormalities are greater when joining regions with abnormal volumes}

\begin{figure}[H]
\centering
\includegraphics[width=1\textwidth]{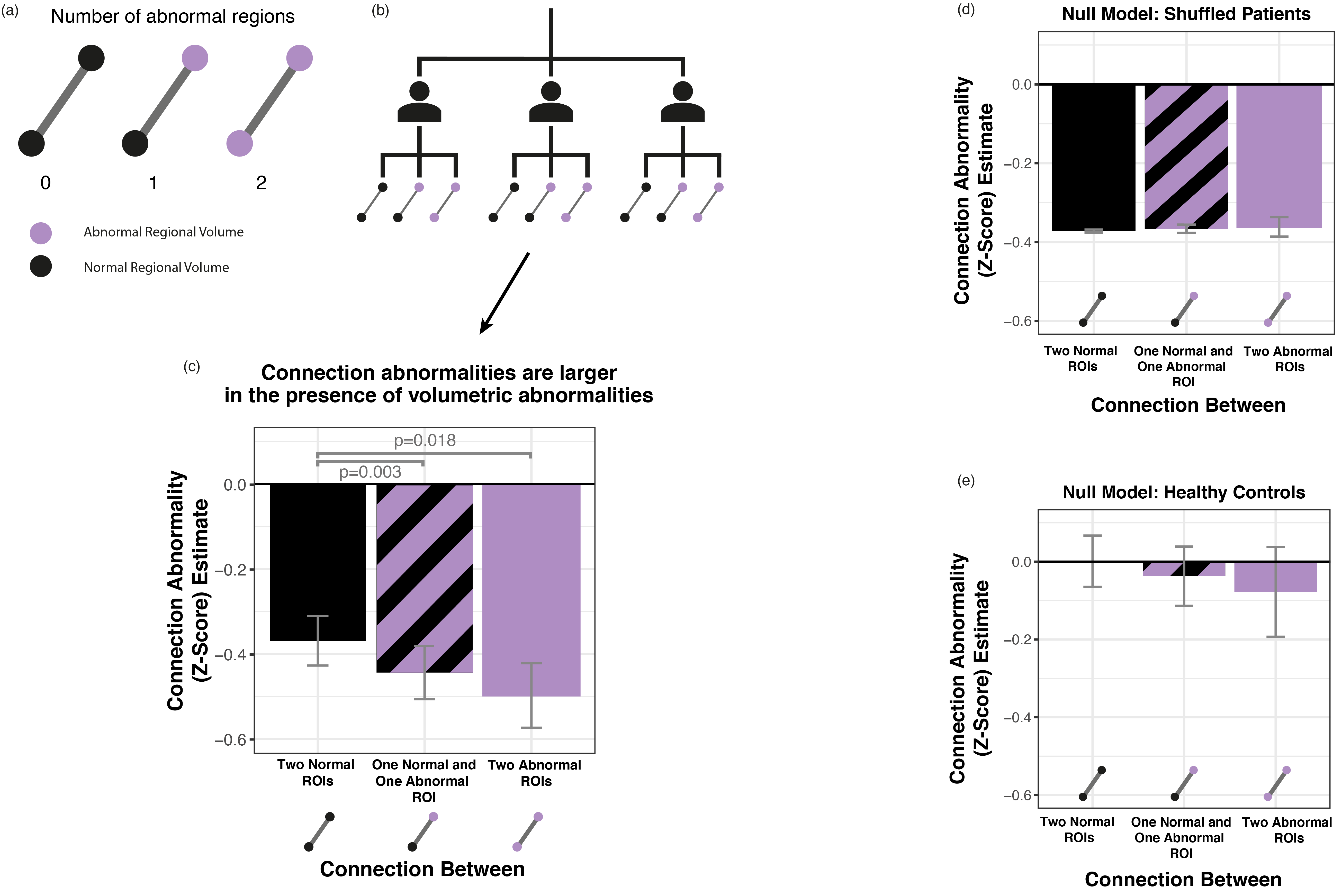}
\caption{\label{fig:Fig2}\textbf{Connection abnormalities are greater when joining regions with abnormal volumes.} a) Each connection within a patient can be connected to zero, one or two abnormal regions. We hypothesised that connections would have greater reductions in FA if they were connected to abnormal regions. b) The individual connection and volumetric abnormalities within each subject formed the input for the hierarchical model. c) Connections had significantly reduced FA when they were connected to one or two abnormal regions. The error bars indicate the standard error of the model estimate. Null models were fitted for comparison using d) patients with the connections shuffled and e) healthy controls.}
\end{figure}

The previous analysis in Figure \ref{fig:Fig1} did not fully account for individual subject-level co-localisations. This is addressed using a hierarchical approach in Figure \ref{fig:Fig2}. Since we used a random intercepts and slopes model, the relationship between abnormalities are modelled individually within each subject, so that the mean connection abnormality between none, one or two abnormal regions is estimated, accounting for random effects. The individual estimates are then aggregated across all subjects to produce overall model estimates of the mean connectivity abnormality in each case. We hypothesised that greater connection abnormalities would coincide with neighbouring volumetric abnormalities. \\

In patients with TLE, connections between one normal and one atrophied region (estimate = -0.446 $\pm$ 0.063; p=0.003), and connections between two atrophied regions (-0.492 $\pm$ 0.075; p=0.018) had significantly reduced FA compared to connections between two normal regions (Figure \ref{fig:Fig2}c). Connections between two normal regions still exhibited reduced FA in comparison to healthy controls (-0.378 $\pm$ 0.059). These results were robust to different $z$-score thresholds (see Supplementary Analysis 4). \\

We fitted two null models for comparison. Firstly, we randomly permuted the connection abnormalities within patients (Figure \ref{fig:Fig2}d). As expected, the mean connection abnormality did not significantly differ depending on whether the connections were joining two normal regions (-0.387 $\pm$ 0.0003), one normal and one abnormal region (-0.382 $\pm$ 0.0008; p=0.53), or two abnormal regions (-0.388 $\pm$ 0.0242; p=0.94). Secondly, we applied the same hierarchical modelling approach to healthy controls (Figure \ref{fig:Fig2}e). Again, the mean connection abnormality did not significantly differ depending on whether the connections were joining two normal regions (0.001 $\pm$ 0.066), one normal and one abnormal region (-0.043 $\pm$ 0.076; p=0.26), or two abnormal regions (-0.089 $\pm$ 0.117; p=0.37).

\subsection{Volumetric abnormalities are greater when joined by abnormal connections}

In complement to the previous section, we hypothesised that regions connected by abnormal connections would be more atrophied than regions connected by normal connections. Specifically, we modelled the mean volume $z$-score of two connected regions, using the thresholded connection abnormality as a binary predictor. \\

\begin{figure}[H]
\centering
\includegraphics[width=1\textwidth]{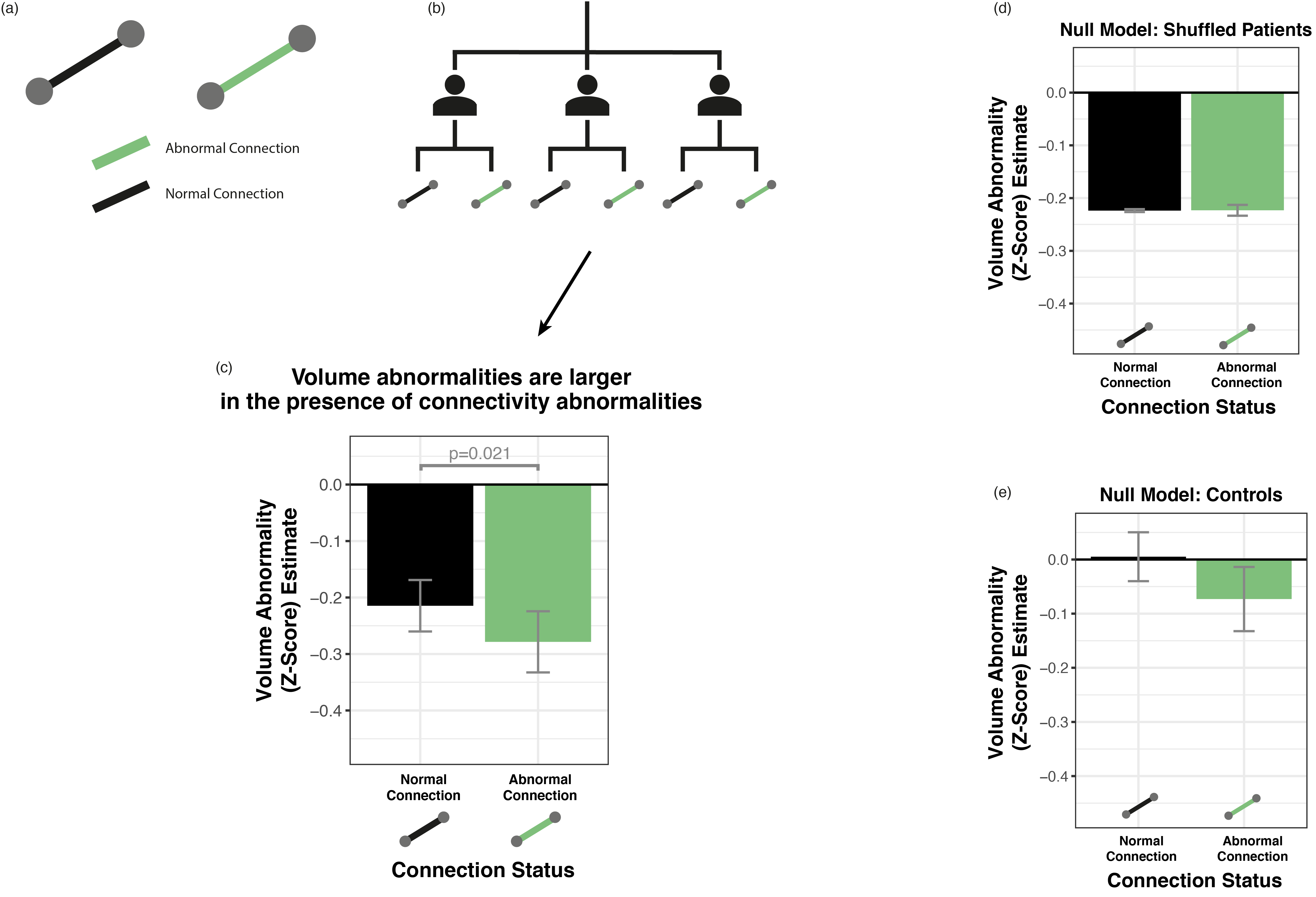}
\caption{\label{fig:Fig3}\textbf{Volumetric abnormalities are greater when joined by abnormal connections.} a) The connection joining two regions can be either normal or abnormal. We hypothesised that the mean volume of two connected regions would have a greater reduction if the regions were joined by an abnormal connection. b) The individual connection and volumetric abnormalities within each subject formed the input for the hierarchical model. c) Regions had significantly reduced volume when they were connected by an abnormal connection. The error bars indicate the standard error of the model estimate. Null models were fitted for comparison using d) patients with the connections shuffled and e) healthy controls.}
\end{figure}

In TLE patients, the volume of regions connected by abnormal connections were significantly reduced, as compared to regions connected by normal connections (estimate = -0.266 $\pm$ 0.052; p=0.021). This is shown in Figure \ref{fig:Fig3}c. The volumes of regions connected by normal connections still exhibited atrophy in comparison to healthy controls (-0.206 $\pm$ 0.045). These results were robust to different threshold values (see Supplementary Analysis 4). \\

Again, we fitted two null models for comparison. Firstly, we randomly permuted the connection abnormalities within patients (Figure \ref{fig:Fig3}d). As expected, the mean volume abnormality did not significantly differ depending on whether regions were connected by a normal (-0.219 $\pm$ 0.046) or abnormal connection (-0.214 $\pm$ 0.056; p=0.51). Secondly, we applied the same hierarchical modelling approach to healthy controls (Figure \ref{fig:Fig3}e). Again, the mean volume abnormality did not significantly differ depending on whether regions were connected by a normal (0.002 $\pm$ 0.045) or abnormal connection (-0.065 $\pm$ 0.058; p=0.07).

\subsection{Volume and connectivity abnormality co-localisation differs across patients}

Next, for each patient, we determined the extent to which abnormalities in the two modalities co-localised. Specifically, we investigated to what extent were atrophied regions found to be connected by abnormal connections. We used Dice similarity as a measure of overlap, then randomly permuted each patients' connection abnormalities 5000 times to generate a null distribution of Dice similarities, given each patients' number of abnormalities. Our measure of co-localisation for each patient was the Dice similarity percentile as compared to the null distribution, and this value could be between zero (not co-localised) and one (co-localised). \\

\begin{figure}[H]
\centering
\includegraphics[width=1\textwidth]{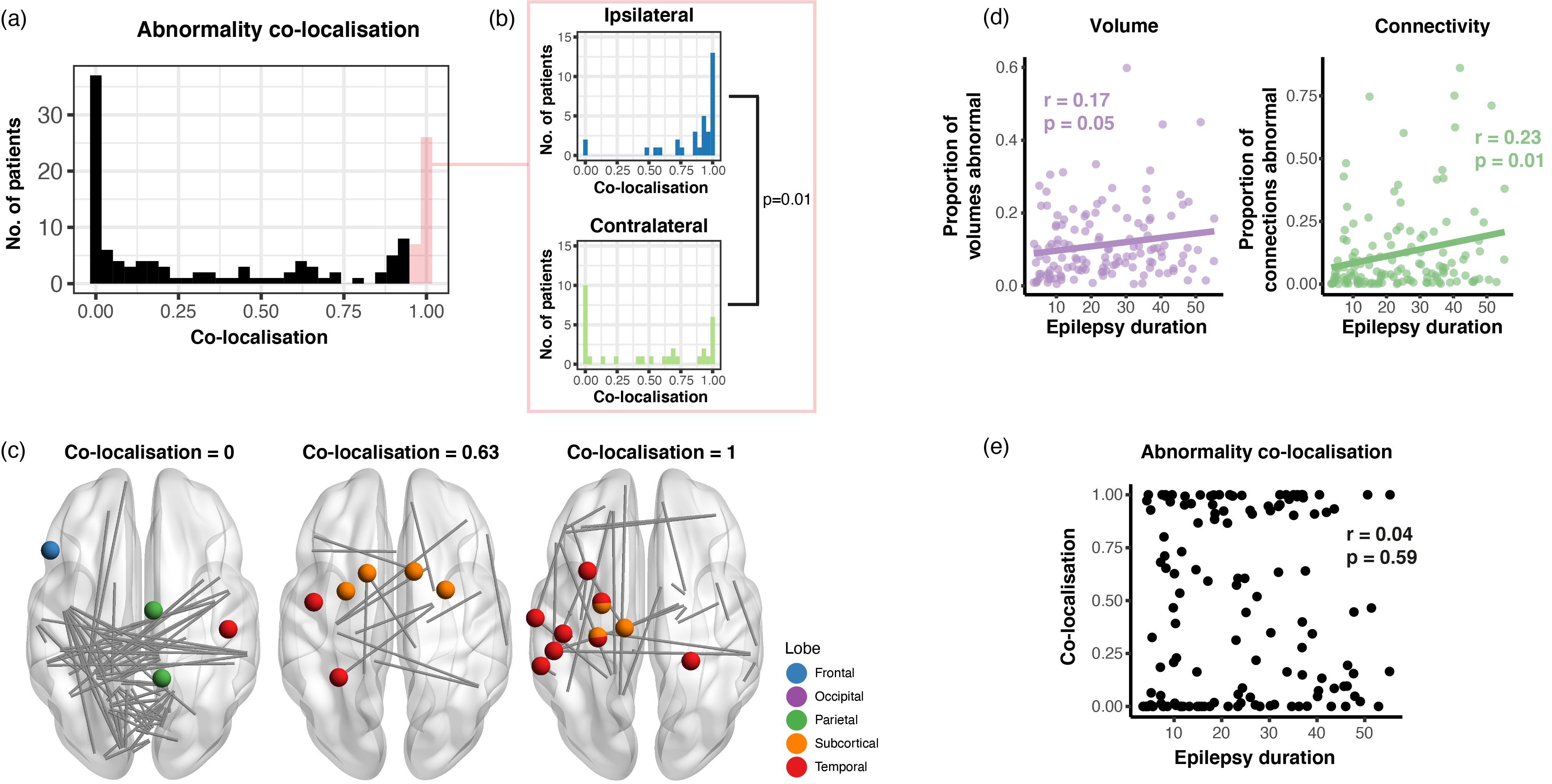}
\caption{\label{fig:Fig4}\textbf{Abnormalities co-localised in some patients, but not others.} a) Abnormalities co-localised in some patients, but not others. b) Of the patients whose abnormalities co-localised (co-localisation $\geq 0.95$), this was significantly more likely to be driven by co-localised abnormalities in the ipsilateral hemisphere. c) Three example patients are given for whose abnormalities were not co-localised, partially co-localised, and co-localised. d) Both the proportion of abnormal regional volumes and abnormal connections increased as epilepsy duration increased. e) However, the co-localisation of abnormalities did not relate to epilepsy duration.}
\end{figure}

Abnormalities co-localised in 30\% of patients, did not co-localise in 40\% of patients, and co-localised to some degree in the remaining patients (Figure \ref{fig:Fig4}a). Of those patients whose abnormalities co-localised (defined as co-localisation $\geq$ 0.95), co-localisation was significantly higher in the ipsilateral hemisphere than the contralateral hemisphere (p=0.01) (Figure \ref{fig:Fig4}b). Three example patients are presented in Figure \ref{fig:Fig4}c with Co-localisation = 0, Co-localisation = 0.63 and Co-localisation = 1. \\

\begin{table}[ht]
\centering
\begin{tabular}{c|c|c|c}
 \hline
  & Co-localisation & Number of Vol & Number of Conn \\ [0.5ex] 
 \hline\hline
 Surgical Outcome & 0.21 (0.20) & -0.06 (0.81) & 0.01 (0.72) \\
 \hline
 Epilepsy Duration & 0.04 (0.59) & \textbf{0.17 (0.05)} & \textbf{0.23 (0.005)} \\
 \hline
 Secondary Generalisation & 0.19 (0.56) & 0.04 (0.87) & -0.23 (0.24) \\
 \hline
\end{tabular}
\caption{Co-localisation differences between groups showing effect size (p-value). Cohen's d and Mann-Whitney U test were used for surgical outcome and secondary generalisation. Pearson correlation coefficient was used for epilepsy duration. Surgical outcome was defined as ILAE 1 vs ILAE 2+.}
\label{table:3}
\end{table}

Both the proportion of regional volumes and the proportion of connections that were abnormal increased as epilepsy duration increased (Pearson's r=0.18, p=0.028; r=0.23, p=0.005 respectively) (Figure \ref{fig:Fig4}d). However, these abnormalities did not necessarily occur in the same regions (i.e. co-localise) and Figure \ref{fig:Fig4}e shows that co-localisation did not increase as epilepsy duration increased (r=0.04, p=0.59). Neither the number of volumetric abnormalities nor the number of connectivity abnormalities within a patient was related to surgical outcome or secondary generalisation of seizures (Table \ref{table:3}).

\section{Discussion}

We investigated the relationship between volumetric and structural connectivity abnormalities in patients with TLE. We first analysed these abnormalities at a group-level, which assessed the location of these abnormalities separately and jointly. Next, we used hierarchical modelling to uncover the relationship between volumetric and structural connectivity abnormalities within individual patients. Finally, we determined if abnormalities co-localised in patients more than would be expected by chance. We found that both volumetric and structural connectivity abnormalities were greater when they coincided with abnormalities in the other modality, and these abnormalities were primarily in ipsilateral subcortical and temporal regions. Volumetric and structural connectivity abnormalities separately were more widespread as epilepsy duration increased, but their co-localisation (i.e. the extent to which they overlap beyond chance, given their prevalence) did not change. These findings suggest that common, or related, mechanisms may underlie changes in both volume and connectivity in patients with TLE.\\

At a group level, our volumetric results were broadly consistent with the literature. Atrophy was unsurprisingly largest in the ipsilateral hippocampus, since 52\% of our patients had hippocampal sclerosis and hippocampal atrophy has been reported frequently in TLE \cite{whelan_structural_2018} \cite{farid_temporal_2012} \cite{keller_morphometric_2015}. We also found reduced volume in the ipsilateral thalamus, as shown in a recent ENIGMA study \cite{whelan_structural_2018}, as well as in the ipsilateral temporal lobe, specifically the temporal pole, middle and superior temporal gyri in agreement with Moran \textit{et al} (2001) \cite{moran_extrahippocampal_2001}. Similarly, our connectivity results concur with previous findings. Connections with the largest FA reductions were connected to bilateral temporal, bilateral subcortical, and ipsilateral frontal regions. The location of these FA reductions strongly aligns with a previous study, which found that connections were most disrupted in temporal and subcortical regions in TLE, albeit to varying degrees \cite{besson_structural_2014}. The largest study of white matter abnormalities investigated specific white matter regions and found the largest FA reductions in ipsilateral external capsule and parahippocampal cingulum \cite{hatton_white_2020}. We instead looked at connections between grey matter regions, and found regions with multiple FA reductions in similar areas of the brain, specifically putamen, pallidum and temporal gyri bilaterally, as well as ipsilateral temporal pole and amygdala. Bilateral reductions in FA of white matter tracts have previously been reported \cite{concha_white-matter_2009}.  \\


Neurobiological explanations for reduced volumes and connectivity have been postulated from experimental and other data. For example, during seizures, excessive glutamatergic neurotransmission can cause grey matter cell bodies to be flooded with calcium leading to excitotoxicity, osmolytic stress and eventually cell death \cite{henshall_apoptosis_2007}. In addition to excitotoxicity, other mechanisms for grey matter cell death exist such as reduced post-ictal localised blood flow (ischaemia/hypoxia) \cite{farrell_neurodegeneration_2017} \cite{farrell_postictal_2016} \cite{fricker_neuronal_2018}, or protein aggregation. The unfolding of these proteins causes protein deposition, triggering degenerative signals in neurons \cite{fricker_neuronal_2018}. In particular, hyperphosphorylation of tau into neurofibrillary tangles has been implicated in TLE \cite{tai_hyperphosphorylated_2016} \cite{gourmaud_alzheimer-like_2020} \cite{zheng_hyperphosphorylated_2014} \cite{alves_tau_2019}. The specific mechanism(s) for atrophy in any given region/patient is unclear, and difficult to determine from MRI (although see e.g. \cite{gaxiola-valdez_seizure_2017} \cite{kim_tau_2020} for recent attempts). However, our imaging findings do strongly suggest that the observed atrophy co-localises with white matter connectivity alterations in at least some individuals. These co-localised connectivity alterations may reflect axonal degeneration following upstream cell body loss. Whilst diffusion MRI-based tractography cannot be used infer the direction of connections, our results demonstrate a greater degree of atrophy in regions joined by connections with reduced FA. These results may reflect the loss of downstream grey matter neurons, although longitudinal studies are needed to confirm causality. \\

Network properties may also influence the relationship between volumetric and structural connectivity abnormalities. Hub nodes are regions with connections to many other regions and are of significance in TLE, since they may spread seizures or pathology far around the network \cite{stam_modern_2014,abdelnour2015relating}. At a group-level, grey matter atrophy co-localises with hub regions \cite{lariviere_network-based_2020} \cite{crossley_hubs_2014}. In patients with generalized tonic-clonic seizures, hub regions have connectivity reductions \cite{li_disrupted_2016}, specifically in connections to other hub regions. The thalamus is known to be a hub region in the brain \cite{hwang_human_2017}, and is particularly important in TLE \cite{caciagli_thalamus_2020}. At a group-level, we found both reduced volume in the ipsilateral thalamus, and connectivity reductions between ipsilateral thalamus and temporal lobe. If hubs are particularly susceptible to both grey matter atrophy and structural connectivity reductions, then the chance of observing both abnormalities \textit{simultaneously} is likely to be increased. Our results were driven by co-localised abnormalities in ipsilateral temporal and subcortical regions, suggesting that TLE has the greatest effect on grey matter and axons in epileptogenic regions, and weaker effects elsewhere. \\

Alongside the observed relationship between volumetric and structural connectivity abnormalities, our model estimated that connections between normal regions still have reduced FA on average, compared to healthy controls (Figure \ref{fig:Fig2}c, solid black bar). Similarly, the volume of regions connected by normal connections are still significantly reduced on average compared to healthy controls (Figure \ref{fig:Fig3}c, solid black bar). These reductions suggest that there is some, but not complete spatial overlap between volumetric and structural connectivity abnormalities within patients. Therefore, some complementary information exists between the two modalities. It could be the case that connectivity abnormalities are more extensive than volumetric abnormalities in some patients, or vice versa, and this will be investigated in future work. \\


Our approach may have clinically relevant implications for two key reasons. Firstly, the incorporation of multiple modalities in computational methods has significant potential to improve localisation of epileptogenic zone \cite{jirsa2017virtual} \cite{proix2017individual}. The epileptogenic zone is the region indispensable for generating seizures \cite{rosenow_presurgical_2001}. In addition to the modalities presented here, we know that multiple modalities separately offer some ability to localise cortical zones that may be related to the epileptogenic zone, including fMRI \cite{morgan_resting_2004}, scalp EEG \cite{sakai_localization_2002} \cite{liang_scalp_2020} \cite{li_localization_2016}, iEEG \cite{taylor_normative_2021} and MEG \cite{nissen_localization_2018} \cite{englot_epileptogenic_2015}. Quantitatively combining modalities could therefore greatly improve our localisation and understanding of the epileptogenic zone, and as a result, improve quality of life for patients with drug-resistant TLE. Secondly, it is important that this localisation of the epileptogenic zone can be done on individual patients. Abnormalities which co-localise spatially across multiple modalities may help to localise the epileptogenic zone \cite{coan_eeg-fmri_2016}. Multiple abnormalities across modalities can be considered simultaneously and quantitatively using methods such as Mahalanobis distance \cite{owen_multivariate_2021} \cite{morgan_mri_2021} or using multilayer or multiplex networks \cite{yu_selective_2017}. Co-localised abnormalities in multiple modalities outside of the surgically resected regions may indicate a failure to remove critical parts of epileptogenic network, leading to poor post-surgical outcomes. \\

A strength of our study is the sample size of 144 patients and 96 healthy controls. Outside multi-centre studies such as the ENIGMA-Epilepsy working group, our sample is a large size for a single centre study. The use of two cohorts of subjects, using different scanning parameters (albeit at the same site) suggests our results are somewhat generalisable. This generalisability could be further improved by the inclusion of additional sites. To our knowledge, our use of a hierarchical modelling approach to examine the relationship between abnormalities across modalities is novel in epilepsy research. \\

One potential limitation of our approach is our heterogeneous cohort of patients at various stages and with potentially different subtypes of temporal lobe epilepsy. The relationship between volume and connectivity abnormalities may differ depending on a patient's stage or subtype, and individual patients may have distinct patterns of gray and white matter structural pathology \cite{lee_decomposing_2021}. For example, it has been suggested that patients with hippocampal sclerosis may have more widespread connectivity abnormalities than those with non-lesional TLE \cite{hatton_white_2020} \cite{bernhardt_hippocampal_2019}, and this may also effect the relationship with volumetric abnormalities. Additionally, in our group-level analysis, we flipped the right and left hemispheres in RTLE patients to perform an ipsilateral-contralateral analysis. This may be important, since it has been acknowledged that right and left hemispheres are not identical \cite{besson_structural_2014}. However, this was done only after z-scoring against the same hemisphere in controls and there is precedence for this flipping in the literature \cite{bonilha_presurgical_2013} \cite{munsell_evaluation_2015} \cite{ji_connectome_2015} \cite{taylor_impact_2018}. Additionally, our hierarchical modelling approach and co-localisation measure are independent of hemisphere (although hemispheres were eventually compared for co-localisation). \\

Distinct neuroimaging modalities offer different perspectives on neurological disorders such as TLE. Considering only one modality may omit useful information for diagnosing and treating these conditions. There is a need to understand how these different modalities highlight different aspects of neurological disorders, separately and jointly. Several other modalities from epilepsy monitoring are available, including time series data (e.g. from EEG, MEG, fMRI). Currently clinicians use these different modalities for diagnosis and localisation of epileptogenic tissue. Developing quantitative methods to improve treatment is vital for improving quality of life for patients. \\

\section{Acknowledgements}

The authors acknowledge the facilities and scientific and technical assistance of the National Imaging Facility, a National Collaborative Research Infrastructure Strategy (NCRIS) capability, at the Centre for Microscopy, Characterisation, and Analysis, the University of Western Australia.
G.P.W. was supported by the MRC (G0802012, MR/M00841X/1). P.N.T. is supported by a UKRI Future Leaders Fellowship
(MR/T04294X/1). J.J.H. is supported by the Centre for Doctoral Training in Cloud Computing for
Big Data (EP/L015358/1).

\printbibliography

\end{document}